\documentclass{article}
\usepackage{spconf,amsmath,graphicx}
\usepackage{amssymb}
\usepackage{amsfonts}
\usepackage{textcomp}
\usepackage{stfloats}
\usepackage{hyperref}       
\usepackage{url}
\usepackage{booktabs}       
\usepackage{verbatim}
\usepackage[utf8]{inputenc}
\usepackage[ruled,vlined,linesnumbered]{algorithm2e}
\usepackage{tikz}
\usepackage{mathtools}
\usepackage{multirow}
\usepackage{caption}
\usepackage{nicefrac}       
\usepackage{microtype}      
\usepackage{xcolor}         
\usepackage{graphicx}
\usepackage{gnuplottex}
\usepackage{wrapfig}

\usepackage{subcaption}

\DeclareMathSymbol{\shortminus}{\mathbin}{AMSa}{"39}

\title{EEG-NeXt: A Modernized ConvNet for the Classification of \newline Cognitive Activity from EEG}
\name{Andac Demir, Iya Khalil, Bulent Kiziltan}
\address{Novartis, Cambridge, USA}
\begin{document}
\ninept
\maketitle
\begin{abstract}
One of the main challenges in electroencephalogram (EEG) based brain-computer interface (BCI) systems is learning the subject/session invariant features to classify cognitive activities within an end-to-end discriminative setting. We propose a novel end-to-end machine learning pipeline, EEG-NeXt, which facilitates transfer learning by: i) aligning the EEG trials from different subjects in the Euclidean-space, ii) tailoring the techniques of deep learning for the scalograms of EEG signals to capture better frequency localization for low-frequency, longer-duration events, and iii) utilizing pretrained ConvNeXt (a modernized ResNet architecture which supersedes state-of-the-art (SOTA) image classification models) as the backbone network via adaptive finetuning. On publicly available datasets (Physionet Sleep Cassette and BNCI2014001) we benchmark our method against SOTA via cross-subject validation and demonstrate improved accuracy in cognitive activity classification along with better generalizability across cohorts.
\end{abstract}
\section{Introduction}
\label{sec:intro}

In order to decode the cognitive activity in the brain, several instruments have been tested to date. MRI analyzes the pulse of radio energy generated from hydrogen nuclei in the tissues upon electromagnetic radiation that penetrates the skull. fMRI detects the presence of oxygen in the blood. PET tracks the energy flow in the brain by tracking the positrons emitted from radioactive glucose, which accumulates within active areas of the brain due to their higher metabolism rate. The greatest advantage of EEG compared to these instruments is its convenience to instantly scan electrical activity in the brain (high temporal resolution), safety, noninvasiveness, portability and low cost. Although EEG has poor spatial resolution, and is vulnerable to artifacts caused by slight motions like eye blinking, which can sometimes render it useless, as well as changes in the data distribution across cohorts and data acquisition sessions, which poses a transfer learning problem, rapid advancements in deep learning, specifically in computer vision, encourage us to develop robust BCI systems that rely on the features extracted from EEG by convolutional neural networks (CNN). These BCI systems present practical solutions in healthcare such as assisting people disabled by neuromuscular disorders~\cite{wolpaw2007brain}, assessing neuropsychiatric diseases~\cite{zhang2021identification}, developing personalized precision psychiatric medicines~\cite{wu2020electroencephalographic}, and augmenting the performance of surgeons~\cite{olivieri2015bci}. 

EEGNet~\cite{lawhern2018eegnet} has been widely adopted in the BCI field as the SOTA method to decode EEG features. A well-known limitation of EEGNet, and similar CNN based architectures, is their inability to explore/exploit the complex functional neural connectivity in the brain. These approaches cannot effectively learn the regional covariates in the brain because EEG data is represented as a pseudo-image where electrodes used during data acquisition are assumed to be equidistant analogous to the pixels of an image. Besides, CNNs learn spatial hierarchies of patterns starting with high frequency features such as corners and edges. However, the information encoded in power spectra of EEGs are dominated by low frequency transients.

\section{Related Work}
Classification of cognitive activity from EEG has been a topic of research in the past few decades. Signal processing techniques such as Independent Component Analysis (ICA)~\cite{viola2010using} and machine learning techniques like Support Vector Machines (SVM)~\cite{guler2007multiclass} and Linear Discriminant Analysis (LDA)~\cite{subasi2010eeg} have been used in conjunction with a feature extraction technique via Power Spectral Density (PSD)~\cite{kim2018effective} or entropy measures~\cite{duan2013differential} to classify EEG signals in the past. MDRM~\cite{6046114} performs classification in the Riemannian space using class covariance matrices. In recent years, CNNs have become a de facto standard for feature extraction and classification of EEG, e.g., EEGNet~\cite{lawhern2018eegnet}, DeepConvNet~\cite{schirrmeister2017deep}, ShallowConvNet~\cite{schirrmeister2017deep} and EEG-Inception~\cite{zhang2021eeg}. These architectures are trained with the temporal data presented as a pseudo-image, where discretized time samples from each channel is arbitrarily (ignoring the coordinates of scalp electrodes) stacked into a single row, or with the spectrograms of EEG signals. Besides, modeling the functional neural connectivity between different EEG electrode sites via graph neural networks (GNN) is an active area of research that requires a more systematic and theoretical analysis~\cite{demir2021eeg}. In order to address the problem of subject/session transfer learning, some studies focus on training conditional variational autoencoders through adversarial disentenglement of subject information from the latent space (A-CVAE)~\cite{ozdenizci2019transfer}. 
Nonetheless, the prevalent method used for subject/session transfer learning in deep learning-based EEG classification is finetuning using new data from the target subject/session that requires a tiresome calibration session for every usage~\cite{krauledat2006reducing}.

\section{Methodology}
Similar to the short-time Fourier transform (STFT), where orthogonal basis functions are a set of sine and cosine waves of different frequency, the wavelet transform also projects the signal onto orthogonal basis functions via time-localized correlation operations. However, the spectrogram produced by STFT violates the Gabor's uncertainty principle, which asserts a fundamental limit on the accuracy of time-frequency localization of a signal~\cite{birchfield2016image}. Specifically, uncertainty principle states that the signal should be multiplied with a windowing function whose width is inversely related to the frequency we are seeking to capture, but STFT splits the signal into several windows of equal length~\cite{birchfield2016image, shoeb2005chapter}. Since STFT subjects all frequencies to the same windowing function with a fixed size, window size will be too narrow for low frequencies and too wide for high frequencies of the EEG signal. The smaller the size of the windows, we know better about where a certain frequency has occurred in the signal, but less about the amplitude of the associated frequency component. Likewise, the larger the size of the windows, we know worse about where a frequency has occurred but better about the amplitude of the associated frequency component. One approach to tackle the problems inherent in the STFT involves the application of wavelet transform.

Wavelet transform yields higher resolution both in temporal and spectral domains by adaptively adjusting the window size based on the frequencies that we are interested to extract from the signal~\cite{addison2005wavelet}. Like the STFT, the wavelet transform can be either discrete (DWT) or continuous (CWT). We apply CWT instead of DWT, and produce the scalogram of EEG trials, which represents the frequency content of the signal as a function of time. DWT is a tree-structured filter bank that convolves and downsamples the signal with the scaling and wavelet functions to produce the wavelet coefficients in lowpass and highpass subbands. The same filtering scheme is iteratively applied to the lowpass subband to produce the wavelet coefficients in lowpass and highpass subbands that spans a narrower frequency band. Upon the application of inverse DWT, the signal can be perfectly reconstructed from its wavelet coefficient. Hence, DWT is well suited to lossless signal compression, denoising and transmission. However CWT has a significant advantage over DWT, when the goal is to produce a fine-grained time-frequency analysis and precisely localize signal transients due to their difference on how to discretize the scale parameter, $a$, as shown in Equation \ref{Eq: mother wavelet}. While DWT uses exponential scales where base is 2: $a=2^i$, $i\in\{1,2,3,\dots\}$, CWT uses finer scales with base less than 2: $a=2^{i/v}$, $v\in\mathbb{Z}$ and $v>1$. 

Let $\psi(t)$ denote a mother wavelet function that is continuous both in time and frequency domain, and $\bar{\psi}(t)$ denote its complex conjugate. Basis functions for the CWT are produced by scaling and shifting the mother wavelet,
\begin{equation}
    \psi_{a,b}(t) = \frac{1}{\sqrt{a}}\bar{\psi}\biggl(\frac{t-b}{a}\biggl)
\label{Eq: mother wavelet}
\end{equation}
where nonnegative $a$ denotes the scaling parameter and determines the frequency of the basis function, while $b$ denotes the translation parameter which determines the location of the basis function. In our model, we use complex valued morlet wavelet (CMOR) as the mother wavelet,
\begin{equation}
    \psi(t) = \frac{1}{\sqrt{\pi B}}\exp\biggl(\frac{-t^2}{B}\biggl)\exp(j2\pi Ct)
\end{equation}
where $B$ denotes the time-decay parameter and $C$ denotes the normalized center frequency, where we define the mother wavelet. $B$ equals the inverse of the variance in the frequency domain. It affects how localized the wavelet is in the frequency domain. As $B$ increases, the wavelet energy gets more concentrated around the center frequency which results in slower decay of the wavelet in the time domain. $C$ has a default value of 1, which leads to good results without the need to search the parameter space.

The CWT of a signal is a 2D array, represented with $X_{w}(a,b)$. Each element in $X_{w}(a,b)$ is the integral of the elementwise product of the signal with the appropriate wavelet function,
\begin{equation}
    X_{w}(a, b) = \frac{1}{\sqrt {a}}\int_{-\infty}^{\infty}\!\! x(t)\psi_{a,b}(t)\,dt
\end{equation}

\begin{figure}[!htbp]
\centering
\includegraphics[scale=0.34]{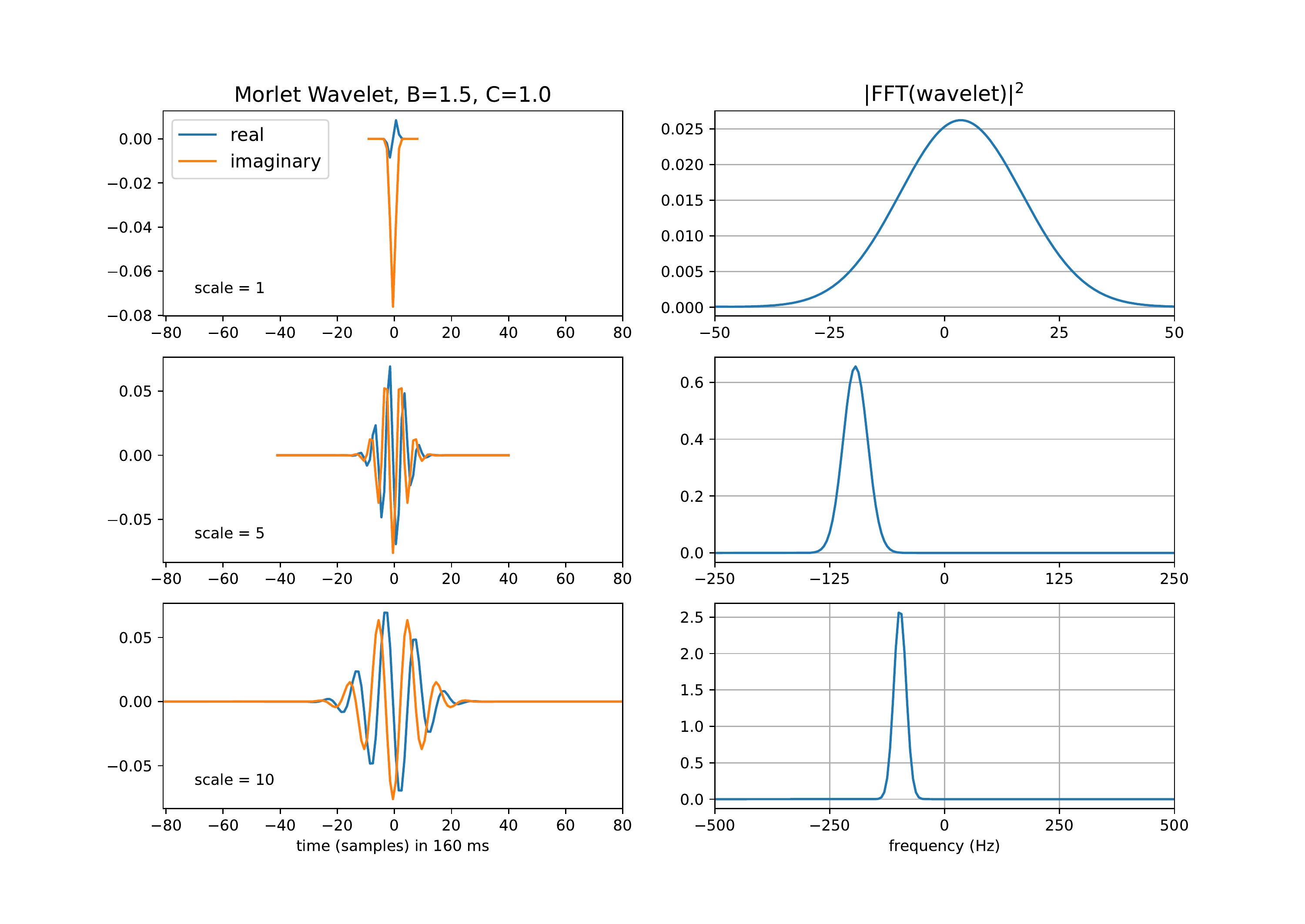}
\caption{\footnotesize The first column shows real (blue) and imaginary (orange) parts of the basis functions produced by CMOR mother wavelet with configuration: $B=1.5$ and $C=1$. The second column shows their corresponding energy spectral density.}
\label{fig:cmor wavelet}
\end{figure}

As illustrated in Figure~\ref{fig:cmor wavelet}, CMOR has the shape of a sinusoid, weighted by a Gaussian kernel, which allows to analyze the event related activity in EEG by capturing local oscillatory components belonging to beta, alpha, theta, and delta bands and gamma waves~\cite{pantazis2005imaging}. Larger values of $a$ corresponds to stretching of the wavelet. For instance, $a=10$ stretches the wavelet by a factor of 10, making it more sensitive to lower frequencies in the signal at the expense of losing temporal resolution. 
\begin{figure}[!htbp]
\centering
\includegraphics[scale=0.5]{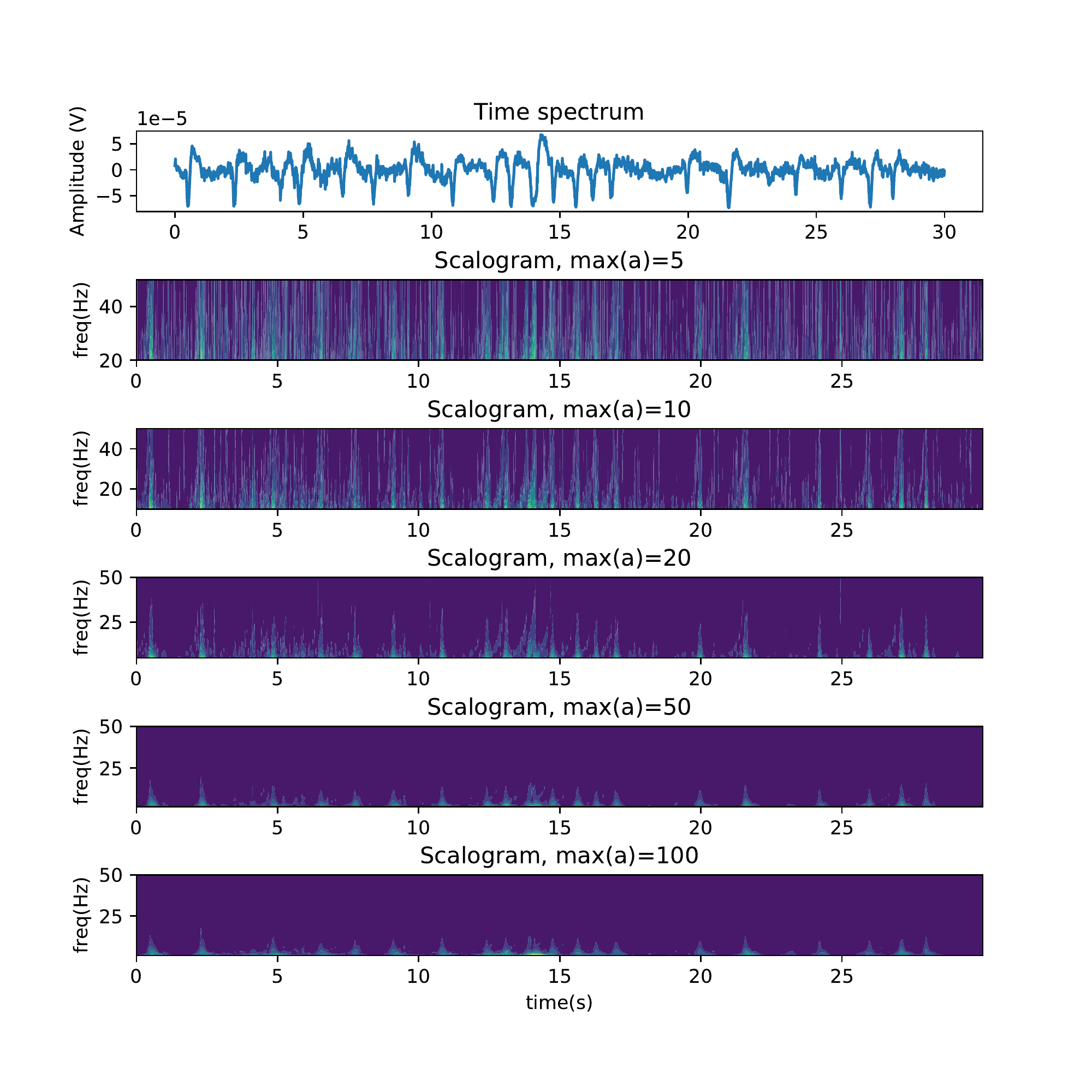}
\caption{\footnotesize Scalograms for $max(a)\in\{5, 10, 20, 50, 100\}$ of an EEG trial from Physionet Sleep Casette Dataset. Given that the EEG signals were sampled at a sampling rate of $f_{s}$, and CMOR was created with a normalized central frequency equal to $C$, scalogram captures the spectral information at $\frac{f_{s}\times C}{a}$ Hz for $\text{scale}=a$.}
\label{fig: scalograms}
\end{figure}

\begin{table*}[t]
\centering
\caption{Implementation details for EEG-NeXt architecture. C denotes the number of channels from which EEG was recorded, S denotes the total number of scales used for CWT computation, T denotes the fixed number of discretized time samples in an EEG trial and L denotes the number of target labels to classify. Number of parameters in the last column is expressed as the summation of number of weight and bias parameters belonging to that layer or a block of layers.}
\label{tab: convnext architecture}
\tiny
\begingroup
\setlength{\tabcolsep}{8pt} 
\renewcommand{\arraystretch}{1.3} 
\begin{tabular}{c c c c c c c c c}
\toprule
\textbf{Layer} & 
\textbf{Filters x (Kernel Size)} &
\textbf{In. features} &
\textbf{Out. features} &
\textbf{Stride} & 
\textbf{Padding} & 
\textbf{Output Dim.} &
\textbf{Options} &
\textbf{Num. of Params.}
\\
\toprule
Input: EEG Scalogram & - & - & - & - & - & (C, S, T) & - & - \\
\cline{1-9}
Conv2D & 3 x (7, 7) & - & - & (1, 1) & (3, 3) &  (3, S, T) & bias=True & 147 + 3\\
\cline{1-9}
GELU & - & - & - & - & - & (3, S, T) & - & - \\
\cline{1-9}
NearestInterpolation & - & - & - & - & - & (3, 64, 64) & - & - \\
\cline{1-9}
Conv2D & 96 x (4, 4) & - & - & (4, 4) & - & (96, 16, 16) & bias=True & 4608 + 96 \\
\cline{1-9}
LayerNorm2D & 96 & - & - & - & - & (96, 16, 16) & eps=1e-6 & 96 + 96 \\
\cline{1-9}
3 x CNBlock-1 & - & - & - & - & - & (96, 16, 16) & - & 237600 \\
\cline{1-9}
LayerNorm2D & 96 & - & - & - & - & (96, 16, 16) & eps=1e-6 & 96 + 96 \\
\cline{1-9}
Conv2D & 192 x (2, 2) & - & - & (2, 2) & - & (192, 8, 8) & bias=True & 73728 + 192 \\
\cline{1-9}
3 x CNBlock-2 & - & - & - & - & - & (192, 8, 8) & -  & 917568  \\
\cline{1-9}
LayerNorm2D & 192 & - & - & - & - & (192, 8, 8) & eps=1e-6 & 192 + 192 \\
\cline{1-9}
Conv2D & 384 x (2, 2) & - & - & (2, 2) & - & (384, 4, 4) & bias=True & 294912 + 384 \\
\cline{1-9}
9 x CNBlock-3 & - & - & - & - & - & (384, 4, 4) & - & 10813824 \\
\cline{1-9}
LayerNorm2D & 384 & - & - & - & - & (384, 4, 4) & eps=1e-6 & - \\
\cline{1-9}
Conv2D & 768 x (2, 2) & - & - & (2, 2) & - & (768, 2, 2) & bias=True & 1179648 + 768 \\
\cline{1-9}
3 x CNBlock-4 & - & - & - & - & - & (768, 2, 2) & - & 14287104 \\
\cline{1-9}
AdaptiveAvgPool2D & 768 & - & - & - & - & (768, 1, 1) & eps=1e-6 & 768 + 768 \\
\cline{1-9}
LayerNorm2D & 768 & - & - & - & - & (768, 1, 1) & eps=1e-6 & 768 + 768 \\
\cline{1-9}
Flatten & - & - & - & - & - & 768 & - & - \\
\cline{1-9}
Linear & - & 768 & L & - & - & L & bias=True & 768 * L + L \\

\bottomrule
\end{tabular}
\endgroup
\end{table*}

\begin{table}
\centering
\caption{Implementation details for CNBlock-1.}
\label{tab: cnblock1}
\tiny
\begingroup
\setlength{\tabcolsep}{0.76pt} 
\renewcommand{\arraystretch}{1.3} 
\begin{tabular}{c c c c c c c c}
\toprule
\textbf{Layer} & 
\textbf{Filters x (Kernel Size)} &
\textbf{In. features} &
\textbf{Out. features} &
\textbf{Stride} & 
\textbf{Padding} & 
\textbf{Options} &
\textbf{\# Params.}
\\
\toprule
Conv2D & 96 x (7, 7) & - & - & (1, 1) & (3, 3) &  bias=True, groups=96 & 4704 + 96 \\
\cline{1-8}
Permute & - & - & - & - & - & (C, H, W) $\rightarrow$ (H, W, C) & - \\
\cline{1-8}
LayerNorm2D & 96 & - & - & - & - & eps=1e-6 & 96 + 96 \\
\cline{1-8}
Linear & - & 96 & 384 & - & - & bias=True & 36864 + 384\\
\cline{1-8}
GELU & - & - & - & - & - & - & - \\
\cline{1-8}
Linear & - & 384 & 96 & - & - & bias=True & 36864 + 96\\
\cline{1-8}
Permute & - & - & - & - & - & (H, W, C) $\rightarrow$ (C, H, W) & - \\ 
\bottomrule
\end{tabular}
\endgroup
\end{table}

\begin{figure}[!htbp]
\centering
\includegraphics[scale=0.385]{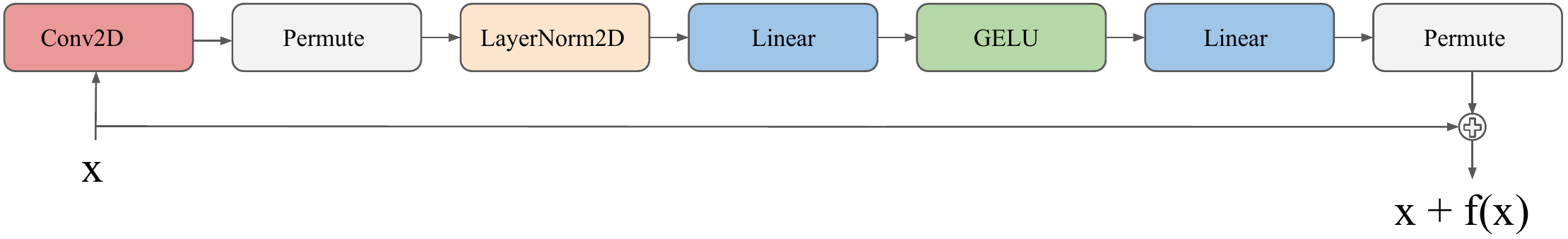}
\caption{\footnotesize Residual connections in CNBlocks.}
\label{fig: convnext block}
\end{figure}

Figure~\ref{fig: scalograms} shows the scalogram of an EEG signal: $X_{w}(a, b)$ for $max(a)\in\{5, 10, 20, 50, 100\}$.  We tweak $a$ to adjust the relative time and frequency resolution of the scalogram. Increasing $a$ gives better frequency resolution, but decreases the time resolution, causes longer edge effects and surges the memory usage of our process~\cite{shoeb2005chapter}. Among a set of candidates, where $max(a)\in\{5, 10, 20, 50, 100\}$, we observe that $max(a)\geq50$ gives significantly better classification performance, but we assign $max(a)=50$ to control memory overhead. This is particularly resulted from the limitation on the frequency resolution of scalogram by the value of $a$. Increasing $max(a)$ increases the frequency resolution of the scalogram, which enables the sliding window technique intrinsic to convolution operation to detect more details and extract fine-grained features from the spectro-temporal analysis of EEG data.    

We finetune ConvNeXt\_tiny~\cite{liu2022convnet} pretrained on  ILSVRC-2012 ImageNet with batches (trials) of multi-channel EEG scalograms. Specifically, we apply 2D convolutions to each EEG trial, which
is presented as a stack of scalograms $\mathbb{R}^{C\times S\times T}$, where $C$ denotes the number of EEG channels, $S$ denotes the total number of scales used for CWT computation and $T$ denotes the number of discretized time samples. This effectively treats the EEG electrode sites and their scalograms like spatial dimensions for CNN processing. 

ConvNeXt improves the ResNet~\cite{he2016deep} architecture by investigating different network architectures and training procedures. ResNet applies to the input $(7\times 7)$ convolutional kernels with stride $(2, 2)$ followed by MaxPool2d. This results in a $(4, 4)$ downsampling, however pooling operation breaks shift-equivariance since high frequency components are aliased as low frequency components. ConvNeXt adopts a ``patchify'' strategy, and eliminates the pooling layer by applying $(4\times 4)$ convolutional kernels with stride $(4, 4)$ to the input image, which effectively yields a $(4, 4)$ downsampling while preserving shift-equivariance. Furthermore, convex and monotonic ReLU activation is replaced with a non-convex and non-monotonic function GeLU. Unlike ReLU, GeLU is differentiable in its entire domain and allows non-zero gradients for negative inputs. Additionally, ConvNeXt uses fewer normalization layers and replaces BatchNorm2d with LayerNorm2d. As LayerNorm2d is independent of batch size, it is effective for training with small batch sizes. Besides, CNBlocks use depthwise convolution instead of standard convolution operations. Depthwise convolution is operated on a per-channel basis and reduces the number of parameters as well as the risk of overfitting. CNBlocks also create an inverted bottleneck, e.g., Table~\ref{tab: cnblock1} shows that the CNBlock takes a low dimensional representation as input, projects it onto high dimension and compresses to low dimension again. Residual connections exist between the low dimensional layers.

\section{Datasets}
\label{section: dataset}
\textbf{Physionet Sleep Cassette Data} is a database~\cite{goldberger2000physiobank, kemp2000analysis} that contains whole-night EEG recordings of 197 subjects.\footnote{ \url{https://www.physionet.org/content/sleep-edfx/}} EEG signals were sampled at 100 Hz. The data were collected in trials of duration 30 seconds using 2 electrodes: Fpz-Cz/Pz-Oz. The data consists of 6 sleep stages: (W)ake, (R)elaxed wakefulness, N1, N2, N3 and REM. EEG trials were labeled based on not only EEG but also on EOG, chin EMG, and event markers by well-trained technicians.

\noindent\textbf{BNCI2014001} is a BCI experiment for motor imagery movement~\cite{tangermann2012review} of the left hand, right hand, both feet and tongue.\footnote{ \url{http://moabb.neurotechx.com/docs/generated/moabb.datasets.BNCI2014001.html\#id1}} 9 subjects were presented with a cue on the screen with the shape of an arrow pointing either to the left, right, down or up (the directions indicate one of the four classes: left hand, right hand, foot or tongue), and instructed to perform the corresponding motor movement. Data was recorded on different days in two sessions for each subject. Each session is composed of 6 runs split by short breaks, and each run includes 48 trials of duration 4 seconds. EEG data were collected using 22 electrodes based on the international 10-10 system with 256 Hz sampling rate. In preprocessing, the signals were bandpass filtered between 0.5 Hz and 100 Hz. An additional notch filter suppresses the line noise at 50 Hz.

\begin{table*}[pt]
\centering
\caption{Benchmark comparison.}
\label{tab: benchmark comparison}

\tiny
\begingroup
\setlength{\tabcolsep}{6pt} 
\renewcommand{\arraystretch}{1.3} 
\begin{tabular}{c c c c c c c c c c c c c c}
\toprule
\textbf{Dataset} & \textbf{Model} &
\multicolumn{6}{c}{\textbf{Classification Accuracy}} & 
\multicolumn{6}{c}{\textbf{ROC-AUC}}\\
\cmidrule(lr){3-8} \cmidrule(lr){9-14}  
& & \textbf{Fold 1} & \textbf{Fold 2} & \textbf{Fold 3} & \textbf{Fold 4} & \textbf{Fold 5} & \textbf{Mean} & \textbf{Fold 1} & \textbf{Fold 2} & \textbf{Fold 3} & \textbf{Fold 4} & \textbf{Fold 5} & \textbf{Mean}\\

\toprule

\multirow{4}{*}{\textbf{Physionet}}
& \textbf{EEG-NeXt (on scalograms)} & 73.7 & 79.5 & 75.7 & 75.7 & 76.1 & \textbf{76.1 \textpm \hspace{0.25em}2.1}& 0.93 & 0.94 & 0.94 & 0.94 & 0.94 & \textbf{0.94 \textpm \hspace{0.25em}0.01}\\
\cline{2-14}

& EEG-NeXt (on temporal data) & 61.0 & 69.7 & 59.1 & 59.9 & 67.5 & 63.4 \textpm \hspace{0.25em}4.8 & 0.87 & 0.91 & 0.88 & 0.87 & 0.89 & 0.88 \textpm \hspace{0.25em}0.02 \\
\cline{2-14}

& EEGNet & 74.0 & 76.1 & 73.4 & 77.5 & 74.5 & 75.1 \textpm \hspace{0.25em}1.7& 0.93 & 0.95 & 0.94 & 0.95 & 0.93 & 0.94 \textpm \hspace{0.25em}0.01 \\
\cline{2-14}

& EEG-Inception & 44.1 & 38.7 & 28.9 & 41.5 & 40.4 & 38.7 \textpm \hspace{0.25em}5.8 & 0.77 & 0.74 & 0.76 & 0.80 & 0.78 & 0.77 \textpm \hspace{0.25em}0.02 \\

\midrule

\multirow{4}{*}{\textbf{BNCI2014001}}
& \textbf{EEG-NeXt (on scalograms)} & 55.5 & 45.9 & 50.2 & 52.5 & 53.2 & \textbf{51.5 \textpm \hspace{0.25em}3.6} & 0.75 & 0.69 & 0.71 & 0.72 & 0.72 & \textbf{0.72 \textpm \hspace{0.25em}0.02} \\
\cline{2-14}

& EEG-NeXt (on temporal data) & 38.0 & 35.3 & 38.6 & 37.2 & 39.0 & 37.6 \textpm \hspace{0.25em}1.5 & 0.61 & 0.57 & 0.63 & 0.61 & 0.62 & 0.61 \textpm \hspace{0.25em}0.02 \\
\cline{2-14}

& EEGNet & 52.8 & 40.9 & 48.1 & 29.3 & 44.4 & 43.1 \textpm \hspace{0.25em}8.9 & 0.74 & 0.66 & 0.67 & 0.52 & 0.66 & 0.65 \textpm \hspace{0.25em}0.08 \\
\cline{2-14}

& EEG-Inception & 37.1 & 37.0 & 39.3 & 39.7 & 37.4 & 38.1 \textpm \hspace{0.25em}1.3 & 0.59 & 0.60 & 0.62 & 0.62 & 0.62 & 0.61 \textpm \hspace{0.25em}0.01 \\
\bottomrule
\end{tabular}
\endgroup
\end{table*}

\section{Experiments}
\label{subsection: Experiments}

\subsection{Euclidean-Space Alignment}
\label{subsection: Euclidean-Space Alignment}
Conventional machine learning algorithms perform poorly in EEG classification due to domain shift in the data distribution between training and test sets. In data preprocessing, we align the EEG trials in the Euclidean space to reduce covariate shift. Let $X_{is}\in \mathbb{R}^{C\times T}$ denote the EEG trial recorded from subject $s$ at trial $i$ and $n$ denote the total number of trials for $s$. First, we compute the  arithmetic mean of covariance matrices from each subject: $\Sigma_{s}=\frac{1}{n}\sum_{i=1}^{n}X_{is}X_{is}^{T}$. Then we transform each EEG trial: $\tilde{X}_{is}=\Sigma_{s}^{-1/2}X_{is}$.

After this transformation, the mean covariance matrix of a subject becomes an identity matrix,
\begin{align}
    \frac{1}{n}\sum_{i=1}^{n}\tilde{X}_{is}\tilde{X}_{is}^{T} &= \frac{1}{n}\sum_{i=1}^{n}\Sigma_{s}^{-1/2}X_{is}X_{is}^{T}\Sigma_{s}^{-1/2} \\
    &= \Sigma_{s}^{-1/2}\biggl(\frac{1}{n}\sum_{i=1}^{n}X_{is}X_{is}^{T}\biggl)\Sigma_{s}^{-1/2} \\
    &= \Sigma_{s}^{-1/2}\Sigma_{s}\Sigma_{s}^{-1/2} = \mathbb{I} 
\label{eq: euclidean space alignment}
\end{align}
If you view the covariance matrices as the feature embeddings of EEG trials, Euclidean-space alignment is similar to the concept of minimizing the maximum mean discrepancy (MMD) across the mean embeddings of subjects.

\subsection{Macro Design}
\label{subsection: Macro Design}

We train our models on a single GPU (NVIDIA Tesla V100 SXM2) by allocating $300$ GB of memory using a total batch size of $128$ for a total of $10$ epochs. We update the model parameters using weighted CrossEntropyLoss as the loss function and AdamW as the optimizer with a constant learning rate $1\mathrm{e}{-4}$, weight decay multiplier $5\mathrm{e}{-4}$, $\beta_{1}=0.9$, $\beta_{2}=0.999$, and $\epsilon=1\mathrm{e}{-8}$. We assess the performance of all models with 5-fold cross-validation. Specifically, the EEG trials collected from one-fifth of subjects are held out for evaluation as a test set and the trials from remaining subjects are used for finetuning ConvNeXt\_tiny. This process is repeated for 5 times, ensuring that the subject data of test sets do not intersect with each other. We use the Python package PyWavelets~\cite{lee2019pywavelets} to perform spectro-temporal analysis of the EEG signals and validate that our results are consistent with results produced from Matlab's wavelet toolbox and functions. 

\subsection{Results}
\label{subsection: Results}
Most of the information of interest in EEG signals is carried by low frequency transients, and scalograms produced via CWT can efficiently reveal long lasting, low frequency oscillations while analyzing nonstationary signals. Table~\ref{tab: benchmark comparison} demonstrates that EEG-NeXt yields promising results on Physionet and BNCI2014001 datasets in cross-subject model transfer scenarios as opposed to two SOTA CNN models: EEGNet and EEG-Inception. We also observe that our performance gain mainly arises from decoding EEG data into a fine-grained spectro-temporal representation rather than employing a more optimized architecture, ConvNeXt, as the backbone network. Superior results seen for scalograms indicate an apparent limitation of CNNs on learning robust features from multivariate time series data represented as a pseudo-image.

In order to find the optimal mother wavelet function, we quantitatively analyze a list of basis functions, specifically CMOR, frequency B-spline (Fbsp) wavelet, Shannon wavelet and complex Gaussian derivative (CGAU) wavelet. Based on our preliminary experiments, the scalograms computed by Fbsp and Shannon wavelets yield sub-par performance for the classification of cognitive activity, whereas the difference between the performances of CMOR and CGAU is negligible.

\section{Conclusion}
Progress in cross-subject/session EEG transfer learning has a significant potential to enhance the scalability of healthcare technologies driven by physiological signals. However the design of neural network architecture is still a big challenge. We proposed a novel end-to-end machine learning pipeline called EEG-NeXt for EEG transfer learning that facilitates learning subject/session invariant features of EEG signals by using scalograms, which provide the spectro-temporal representation of EEG signals and enable us to better capture the low frequency signal transients. We benchmark the performance of EEG-NeXt on Physionet Sleep Cassette and BNCI2014001 datasets. Constructed solely from the ConvNeXt architecture, EEG-NeXt competes favorably against EEGNet and EEG-Inception in terms of classification accuracy of the cognitive activity. The extent to which it is possible to improve the classification performance of three major BCI paradigms, mainly motor imagery, steady-state visual evoked potentials (SSVEP) and event related potential (ERP), requires a more comprehensive experimental study that comprises more datasets.



\vfill\pagebreak


\bibliographystyle{IEEEbib}
\bibliography{refs}

\end{document}